\documentclass[pdflatex,sn-vancouver-num]{sn-jnl}

\usepackage[utf8]{inputenc}
\usepackage{graphicx}
\usepackage{amsmath,amssymb,amsfonts}
\usepackage{amsthm}
\usepackage{mathrsfs}
\makeatletter
\input{ursfs.fd}
\DeclareFontShape{U}{rsfs}{m}{n}{
  <-6> rsfs5
  <6-8> rsfs7
  <8-> rsfs10
}{}
\makeatother
\usepackage[title]{appendix}
\usepackage{xcolor}
\usepackage{booktabs}
\usepackage{bookmark}
\usepackage{siunitx}

\theoremstyle{thmstyleone}

\theoremstyle{thmstyletwo}

\theoremstyle{thmstylethree}

\begin{document}

\title{Efficient Ion-Photon Quantum Interface with Long-Working-Distance Exceeding 26 mm}

\author[1,2]{\fnm{Jungwoo} \sur{Choi}}\email{patrick0308@g.skku.edu}

\author[1,2]{\fnm{Inhyuk} \sur{Oh}}\email{oig0605@g.skku.edu}
\author[1,2]{\fnm{Chanyang} \sur{Im}}\email{cksdid636@g.skku.edu}
\author[2]{\fnm{Gibeom} \sur{Son}}\email{gibeom.son@g.skku.edu}

\author*[1,2]{\fnm{Junki} \sur{Kim}}\email{junki.kim.q@skku.edu}

\affil[1]{\orgdiv{Department of Quantum Information Engineering}, \orgname{Sungkyunkwan University}, \orgaddress{\street{Seobu-ro 2066}, \city{Suwon}, \postcode{16419}, \state{Gyeonggi-do}, \country{Korea}}}

\affil[2]{\orgdiv{Department of Nano Science and Technology \& SKKU Advanced Institute of Nanotechnology (SAINT)}, \orgname{Sungkyunkwan University}, \orgaddress{\street{Seobu-ro 2066}, \city{Suwon}, \postcode{16419}, \state{Gyeonggi-do}, \country{Korea}}}

\abstract{
    Trapped-ion quantum networks rely on efficient ion-photon interfaces, while objective-based free-space collection systems must balance high photon collection with sufficient working distance for practical integration with ion-trap hardware.
    Here, we design and characterize an ion-photon interface with a long working distance of \qty{26.15}{\milli\meter} and a high numerical aperture of 0.5.
    The core of the interface is a photon-collection objective consisting of six commercial spherical lenses with optimized spacing.
    The interface achieved an overall fiber-coupled photon-detection efficiency of \((1.29\pm0.09)\%\) from the ion to the detector and maintains fiber coupling over several hours without active realignment, supporting stable operation of the ion-to-fiber interface. Photon-correlation measurements show clear antibunching in the fiber-coupled fluorescence, confirming the preservation of non-classical photon statistics. These results demonstrate a practical implementation of a long-working-distance photon collection interface for trapped-ion quantum network experiments.
}


\maketitle
\section{Introduction}\label{sec1}

Trapped ions are a leading platform for quantum information processing, providing both high-fidelity quantum control and long-lived internal states~\cite{ballance_high-fidelity_2016, gaebler_high-fidelity_2016, wang_single_2021}.
As trapped ion processors scale to larger qubit numbers, however, the control hardware complexity grows rapidly, making it increasingly difficult to support large registers within a single trap.
The photonic-based modular architecture approach addresses this challenge by linking smaller ion registers through photonic channels~\cite{duan_colloquium_2010, monroe_large_2014, hucul_modular_2015, brown_co-designing_2016, main_distributed_2025}.

In such a modular architecture, the key element is a quantum interface that generates ion-photon entanglement between a stationary ion qubit and an emitted photon~\cite{blinov_observation_2004}.
The ion-entangled photons can then be brought together and made to interfere, enabling remote ion-ion entanglement~\cite{simon_robust_2003, moehring_entanglement_2007, stephenson_high-rate_2020} and quantum teleportation between distant matter qubits~\cite{olmschenk_quantum_2009}.
For this process, the photon must be delivered in a well-defined spectral, spatial, and polarization mode~\cite{stephenson_high-rate_2020, oreilly_fast_2024} so that photons emitted from independent ions can be made indistinguishable at the interference stage.
Coupling the collected fluorescence into a single-mode fiber (SMF) is therefore commonly used to satisfy these criteria, because the SMF defines a well-defined photonic mode and also supports photon delivery between remote nodes~\cite{kim_efficient_2011}.

In addition to mode definition, an ion-photon quantum interface must collect photons with high efficiency.
A larger collection solid angle increases the probability that an emitted photon is captured and coupled into the SMF, which is important for probabilistic photon-mediated entanglement protocols~\cite{monroe_large_2014}. High photon collection also benefits qubit readout via state-dependent fluorescence~\cite{myerson_high-fidelity_2008, noek_high_2013}.
This requirement naturally favors high-numerical aperture (NA) objective lenses. However, increasing the collection NA generally reduces the available working distance (WD) and makes the system more sensitive to aberrations, alignment errors, and mechanical constraints.
For trapped ion experiments, the working distance is a critical design parameter because the collection optics must be compatible with the trap electrodes and surrounding vacuum chamber.
Therefore, a practical objective-based interface should maintain a large collection NA while preserving sufficient working distance for system integration.

Efficient ion-photon interfaces have been demonstrated using high-collection-NA objective lenses with NA as high as 0.8~\cite{carter_ion_2024}, supporting high-rate, high-fidelity remote entanglement~\cite{oreilly_fast_2024, saha_high-fidelity_2025} and distributed quantum-computing demonstrations~\cite{main_distributed_2025}.
However, the short working distance of such optics often complicates ion-trap apparatus design, requiring solutions such as re-entrant viewports~\cite{maunz_quantum_2007, crain_individual_2014} or even an in-vacuum integration of the objective lens~\cite{carter_ion_2024}.
Conversely, several studies have demonstrated custom long-working-distance objectives based on commercial off-the-shelf lenses in single-atom experiments, but with more modest collection NA values up to 0.4~\cite{pritchard_long_2016, li_high-numerical-aperture_2020}.
It is also worth noting that cavity-enhanced interfaces have been proposed and studied to increase photon extraction and ion-photon coupling~\cite{schupp_interface_2021}. However, they require integrating an optical cavity with the ion-trap apparatus, which can introduce technical constraints beyond those of a lens-based free-space collection geometry.

Here, we built and characterized an ion-photon quantum interface with very long working distance and high photon collection efficiency. The custom-designed objective lens provides a \qty{26.15}{\milli\meter} working distance while maintaining an object-space NA of 0.5. We observed a peak fiber-coupling efficiency of \((41.9\pm3.0)\%\), resulting in overall fiber-coupled photon-detection efficiency of \((1.29\pm0.09)\%\). The interface shows resilience against system drift, maintaining a high coupling efficiency for several hours, and exhibits photon antibunching with $g^{(2)}(\tau=0) = 0.042$, indicating non-classical photon statistics.

\section{Experimental Setup}\label{sec:experimental-setup}

\begin{figure}[htbp]
    \centering
    \includegraphics[width=0.9\textwidth]{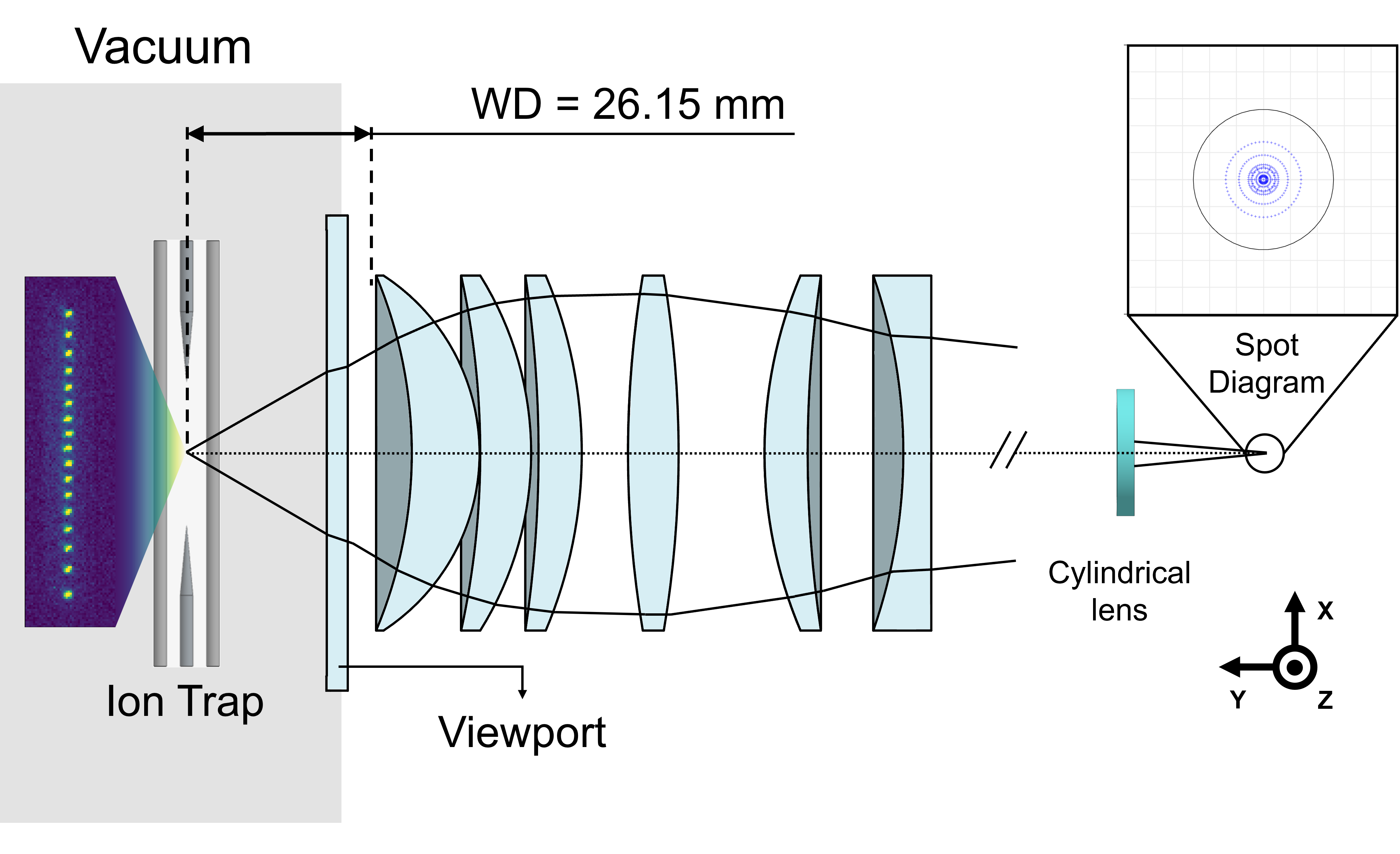}
    \caption{
        Schematic cross-section of the long-WD objective lens system. The working distance of the custom objective is \qty{26.15}{\milli\meter}, defined as the mechanical clearance from the ion position to the assembled objective housing. The gray region marks the vacuum-chamber interior, which houses a four-rod Paul trap that confines the ions (camera image: 17 $^{138}\mathrm{Ba}^{+}$ ions). Black solid lines indicate the marginal rays that define the object-space NA of 0.5. At the focal plane, the full marginal-ray bundle falls within the Airy disk (right inset), as required for the diffraction-limited design. The final cylindrical lens (Thorlabs LJ1516RM-A) is used in the experiment to correct residual astigmatism.
        }\label{fig1}
    \end{figure}

\begin{table}[t]
\centering
\caption{
The lens prescription for the long-WD objective lens system.
\(R_1\) and \(R_2\) are the radii of curvature of the first and second surfaces, respectively, following the sign convention along the propagation direction from the ion to the image plane. All dimensions are in mm.
}
\label{lens data}

\small
\setlength{\tabcolsep}{4pt}

\begin{tabular}{@{}llcccc@{}}
\toprule
Element & Part number & \(R_1\) & \(R_2\) &
\shortstack{Center\\thickness} &
\shortstack{Distance after\\element} \\
\midrule
Object plane & -- & -- & -- & -- & 20.089 \\
Viewport & -- & \(\infty\) & \(\infty\) & 3.000 & 8.511 \\
L1 & LE1076-A & -65.800 & -30.340 & 9.700 & 0.100 \\
L2 & LE1418-A & -119.320 & -47.870 & 7.290 & 1.058 \\
L3 & LE1015-A & -171.600 & -65.160 & 6.180 & 6.624 \\
L4 & LB1374-A & 153.280 & -153.280 & 7.240 & 12.297 \\
L5 & LE1015-A & 65.160 & 171.600 & 6.180 & 13.690 \\
L6 & LC1611-A & -77.190 & \(\infty\) & 4.000 & 141.725 \\
Image plane & -- & -- & -- & -- & -- \\
\bottomrule
\end{tabular}
\end{table}

The ion-photon interface is built around the fluorescence collection objective lens that collects photons emitted from a single trapped ion and couples them into a SMF. The objective lens was designed to satisfy three main requirements: a long WD, a high NA, and diffraction-limited imaging performance.
The required WD was determined by the geometry of the vacuum chamber and viewport. The ion trap is housed in a 4.5-inch spherical octagon chamber equipped with a standard ConFlat (CF) viewport, which requires the WD to exceed \qty{24.5}{\milli\meter}. We targeted an object-space NA of 0.5, corresponding to the collection of 6.7\% of the full solid angle of the ion fluorescence. In addition to efficient photon collection, the objective lens was designed to provide diffraction-limited imaging at the operating wavelength of \qty{493}{\nano\meter}. Since the collected photons are coupled into an SMF, the image-space NA and focal spot size of the objective lens were designed to match the NA and mode-field diameter (MFD) of the fiber (Thorlabs P1-488PM-FC-2). The SMF acts as a spatial-mode filter and projects the collected fluorescence onto a well-defined fiber mode. Such mode definition is important for photon delivery between nodes and for interference-based entanglement swapping protocols, where the spatial overlap of photons from independent ions affects the interference visibility~\cite{kim_efficient_2011}.

Based on these WD, NA, and fiber-mode-matching requirements, we designed, simulated, and optimized the objective lens configuration in Ansys Zemax OpticStudio. The resulting optical layout is shown in Fig.~\ref{fig1}, and the lens element specifications are listed in Table~\ref{lens data}. The configuration comprises six commercial spherical lenses with a 2-inch diameter, with spacings optimized for operation at \qty{493}{\nano\meter}. Accounting for the lens housing, the design yields a working distance of \qty{26.15}{\milli\meter} while the center of the first lens surface is \qty{31.60}{\milli\meter} from the ion along the optical axis.
Ray-tracing simulations confirmed that the system achieves diffraction-limited performance, with a calculated root-mean-square (RMS) spot radius of \qty{1.544}{\micro\meter}, which is smaller than the Airy disk radius of \qty{2.604}{\micro\meter}. The maximum SMF coupling efficiency predicted by the Zemax model, calculated using an overlap integral between the simulated focal field and the fiber mode, was 73.9\%. This value represents the maximum expected under the simulated design conditions and does not include additional aberrations or mode mismatch introduced in the assembled experimental system. A cylindrical lens was additionally used in the experiment to compensate residual astigmatism, as discussed in the following section.

Realizing this design in practice, the relative spacing between adjacent lenses during assembly was critical to achieving the target performance.
We used a Thorlabs SM2 lens tube to hold the six lenses, with each lens held in place by a pair of retaining rings (Thorlabs SM2RR).
From the curved lens geometry and the measured thicknesses of the retaining rings, we calculated the required ring spacing and carefully adjusted the ring positions to match it.
To avoid off-axis alignment errors between lenses, we also kept the lens tube in a fixed orientation throughout the assembly.
Based on the thread pitch and the precision of our spacing measurements, we estimate the resulting error in ring spacing to be below \qty{0.1}{\milli\meter}.
Despite these efforts, we observed shot-to-shot variation in the lens assembly quality.
We built three nominally identical assemblies, and the measured fiber coupling efficiency varied by approximately a factor of three.
In the remainder of the paper, our analysis is based on the assembly with the highest fiber coupling efficiency.

The objective lens assembly was integrated into an ion trap apparatus for fluorescence collection from $^{138}\mathrm{Ba}^{+}$ ions. A linear 4-rod Paul trap confines the barium ions in the vacuum chamber, and the four trap electrodes are arranged in a rectangular configuration to maximize optical access, providing an NA of 0.58 along the optical axis. The vacuum chamber, objective lens, and optics for photoionization and laser cooling were mounted on a common custom-machined base to maintain mechanical stability throughout the measurements.

\section{Results and Discussion}\label{sec3}

Precise alignment of the objective lens is a prerequisite for efficient photon collection via the quantum interface. Measured through-focus images of the ion were compared with ray-tracing simulation results to diagnose and correct optical aberrations, including spherical aberration, coma, and astigmatism. A five-axis lens holder enabled fine alignment to minimize these aberrations, while a cylindrical lens was inserted into the optical path to correct residual astigmatism.

\begin{figure}[htbp]
\centering
\includegraphics[width=0.9\textwidth]{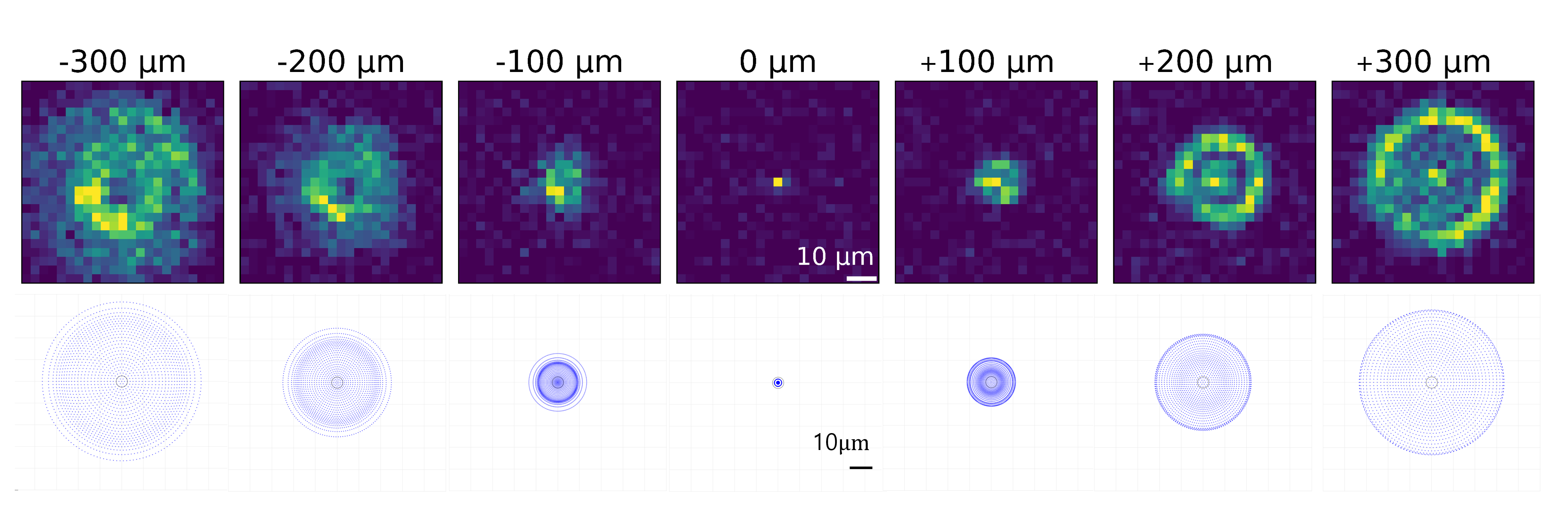}
\caption{Through-focus fluorescence imaging and ray-tracing simulations of a single ${}^{138}\text{Ba}^+$ ion. Experimental images after alignment optimization are shown in the top row, and corresponding ray-tracing simulations are shown in the bottom row. Images were acquired at \qty{100}{\micro\meter} axial steps from \qty{-300}{\micro\meter} to +\qty{300}{\micro\meter} relative to the focal plane. Scale bar: \qty{10}{\micro\meter}.}\label{fig2}
\end{figure}

Fig.~\ref{fig2} shows optimized through-focus images and simulations of the ion obtained by scanning the imaging sensor position along the optical axis from \qty{-300}{\micro\meter} to +\qty{300}{\micro\meter} in \qty{100}{\micro\meter} increments. Given the Airy radius, we expected the fluorescence signal to be concentrated within a single \qty{3.45}{\micro\meter} pixel. Measurements at the focal plane verified this, demonstrating that the fluorescence was indeed confined to a single pixel. The intensity distributions of through-focus images agree well with the ray-tracing simulation results across the scanned range. These observations are consistent with near-diffraction-limited performance, although the complementary metal-oxide-semiconductor (CMOS) camera pixel size prevents a direct quantitative measurement of the focal-plane point-spread function (PSF).
The isotropic structure of the out-of-focus images confirms that coma and astigmatism were sufficiently suppressed, although the residual higher-order aberrations that were not fully corrected by the lens alignment process may remain.

Following the camera-based alignment shown in Fig.~\ref{fig2}, a SMF was positioned at the image focus, and its position was then adjusted to maximize the coupled fluorescence signal.
The fiber-coupling efficiency, $\eta_{\mathrm{fiber}}$, is defined as the ratio between the fiber-coupled photon count rate and the free-space photon count rate measured under the same ion and laser conditions.
After this optimization, the maximum fiber-coupling efficiency was $\eta_{\mathrm{fiber}}=(41.9\pm 3.0)\%$ at a moderate driving-field strength.
When the ion was driven by a stronger field, the fiber-coupled photon count rate reached $312.7 \pm 5.1$~\si{\kilo\hertz}, with a slightly lower coupling efficiency of $32.4\pm1.4\%$.
The reduction is likely due to reduced ion stability under the stronger driving field.
The overall photon detection efficiency through the single-mode fiber (SMF), $\eta_{\mathrm{total}}$, is given by the product of the free-space photon detection efficiency, $\eta_{\mathrm{fs}}$, and the fiber-coupling efficiency, $\eta_{\mathrm{fiber}}$.
Here, $\eta_{\mathrm{fs}}$ accounts for the photon-collection efficiency of the lens system, the optical transmission efficiency, and the detector quantum efficiency.
We obtained $\eta_{\mathrm{fs}}$ by comparing free-space fluorescence measurements with the steady-state population predicted by a three-level optical Bloch equation model, which yielded $\eta_{\mathrm{fs}}=(3.09\pm 0.04)\%$ (see Appendix~\ref{secA1}).
Combining this value with the peak fiber-coupling efficiency gives a best measured total fiber-coupled detection efficiency of $\eta_{\mathrm{total}}=\eta_{\mathrm{fs}}\eta_{\mathrm{fiber}} = (1.29\pm 0.09)\%$.

\begin{figure}[htbp]
\centering
\includegraphics[width=0.9\textwidth]{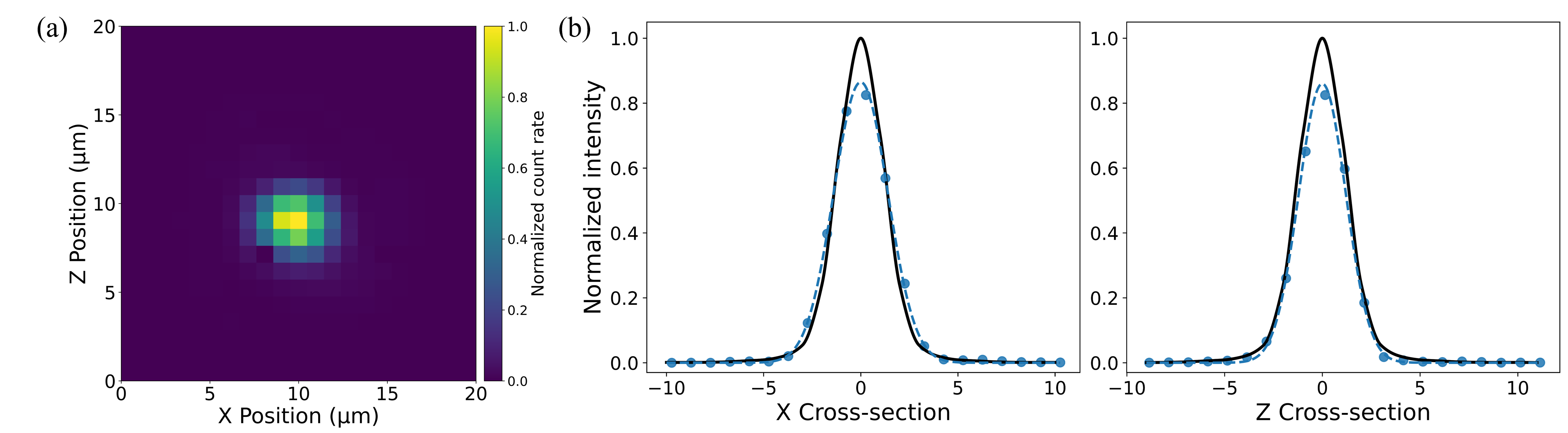}
\caption{Intensity profile characterization by scanning the single-mode fiber tip. (a) Normalized photon count rate over a \qty{20}{\micro\meter} $\times$ \qty{20}{\micro\meter} area (\qty{1}{\micro\meter} steps). (b) Cross-sections along X and Z axes. Experimental data and Gaussian fits (blue markers and blue dashed curves) are compared with the convolution of the simulated PSF and the fiber mode profile (black solid curves).} \label{fig3}
\end{figure}

Further characterization of the fluorescence profile was performed by scanning the fiber tip at \qty{1}{\micro\meter} intervals over a \qty{20}{\micro\meter} $\times$ \qty{20}{\micro\meter} area, as shown in Fig. \ref{fig3}(a). The measured scan data were compared with the convolution of the simulated PSF and the expected fiber mode profile. As shown in Fig. \ref{fig3}(b), the measured $1/e^2$ beam radii of \qty{2.8}{\micro\meter} and \qty{2.4}{\micro\meter} are in close agreement with the \qty{2.3}{\micro\meter} radius of the simulation. This agreement indicates that the overall focal spot size is close to the simulated result, although the fiber-scan measurement alone does not fully characterize the focal-field phase or higher-order aberrations. The measured peak coupling efficiency is lower than the maximum value of 73.9\% predicted by the Zemax model. The remaining difference may arise from additional aberrations introduced in the assembled optical system, uncertainty in the actual fiber mode-field diameter, finite alignment precision, and slow drift of the ion image relative to the fiber mode during fluorescence measurements. These effects are not fully captured in the idealized ray-tracing and overlap integral calculation.

\begin{figure}[htbp]
\centering
\includegraphics[width=0.9\textwidth]{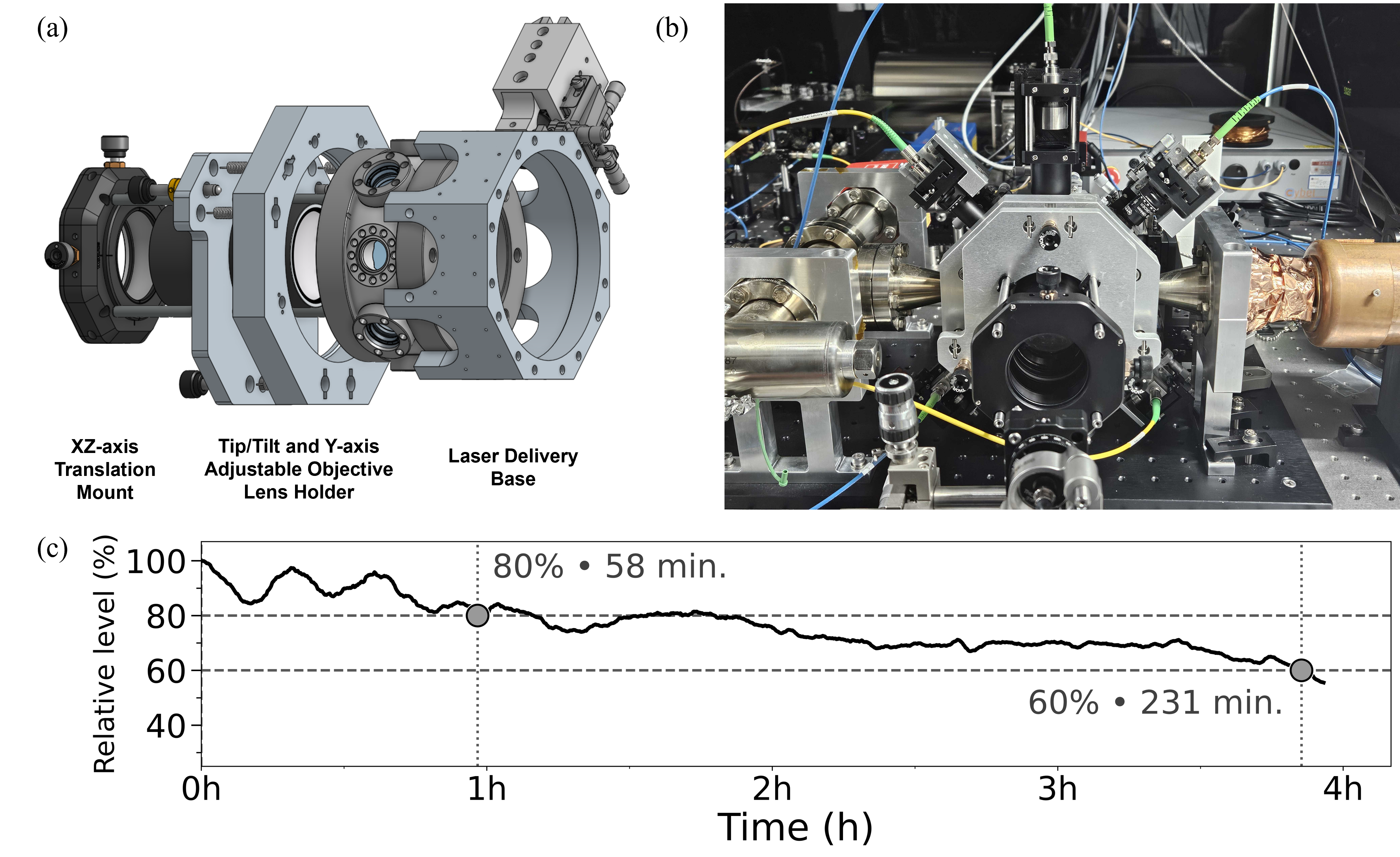}
\caption{Mechanical integration and passive mechanical stability. (a) Computer-aided design (CAD) model of the custom chamber base and objective mount. (b) Photograph of the experimental apparatus. (c) Normalized fiber-coupled fluorescence monitored for \qty{4}{\hour} without realignment, showing the signal remains above 80\% of its initial value for \qty{58}{\minute} and reaches 60\% at \qty{231}{\minute}.} \label{fig4}
\end{figure}

The custom-machined chamber base and objective lens mounting structure are shown in Fig. \ref{fig4}(a,b). The chamber base provides a rigid platform for both the vacuum chamber and the objective lens assembly, helping to reduce relative mechanical motion between the ion trap electrodes and the collection optics. The passive stability of the ion-to-fiber coupling was evaluated by monitoring the fiber-coupled fluorescence signal over a \qty{4}{\hour} period without active realignment. As shown in Fig. \ref{fig4}(c), the signal remained above $80\%$ of its initial value for \qty{58}{\minute} and reached $60\%$ after \qty{231}{\minute}. During the stability measurement, the apparatus was operated in a normal laboratory environment with the building heating, ventilation, and air-conditioning (HVAC) system running and without additional temperature stabilization. The ambient-temperature variation observed by a sensor placed near the chamber remained within about \qty{1}{\celsius} peak-to-peak.

\begin{figure}[htbp]
\centering
\includegraphics[width=0.9\textwidth]{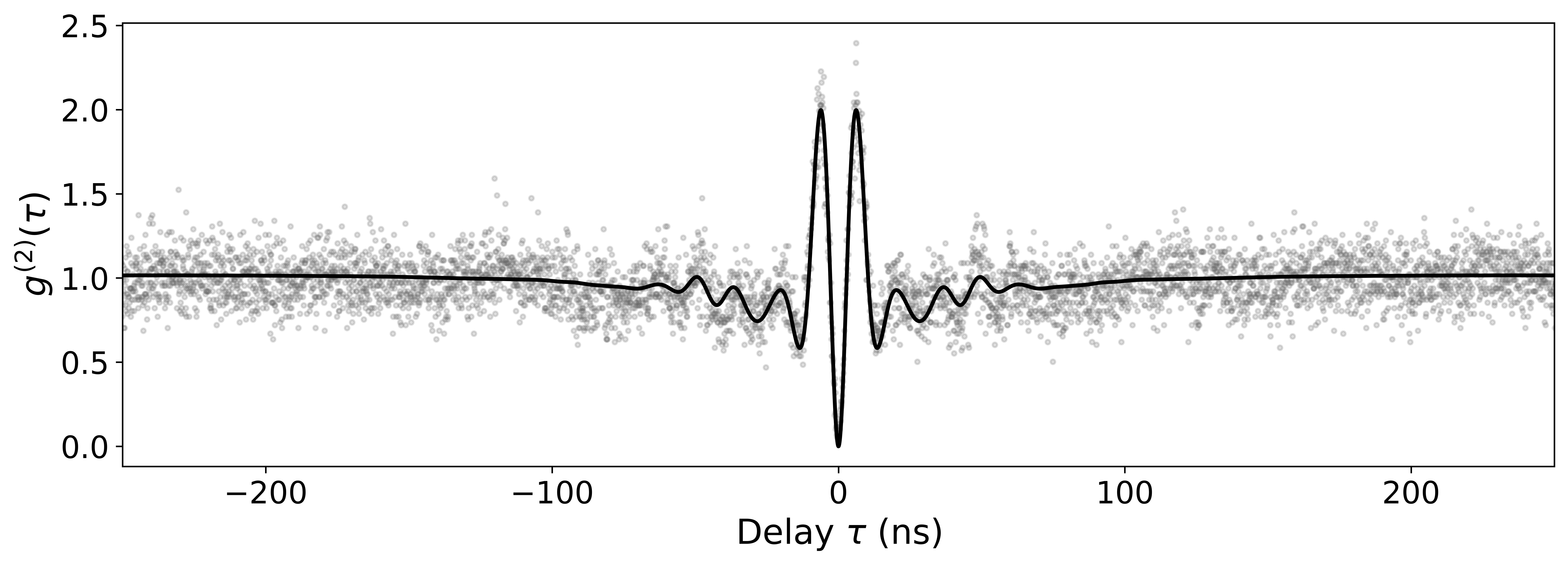}
\caption{Second-order correlation function $g^{(2)}(\tau)$ of single-ion fluorescence coupled to a SMF. Gray dots represent the experimental data, and the black solid line indicates the theoretical fit based on the 3-level optical Bloch equations.}\label{fig5}
\end{figure}

The photon statistics of the collected fluorescence were measured using a Hanbury Brown and Twiss setup \cite{hanbury_brown_test_1956,diedrich_nonclassical_1987}. During the measurement, the ion was continuously Doppler cooled, and the \qty{493}{\nano\meter} fluorescence scattered on the $6S_{1/2} \leftrightarrow 6P_{1/2}$ transition was collected through the objective lens and sent to a 50:50 fiber beam splitter.
The resulting second-order correlation function \(g^{(2)}(\tau)\) shows a suppression of photon coincidences at zero time delay, yielding \(g^{(2)}(0)=0.042\), as shown in Fig. \ref{fig5}. The \(g^{(2)}(\tau)\) data were analyzed using a time-dependent optical Bloch equation model for a three-level \(\Lambda\)-system, where the recovery of the excited-state population after photon detection was calculated and normalized by its steady-state value, \(g^{(2)}(\tau)=\rho_{ee}(\tau)/\rho_{ee}(\infty)\)~\cite{mcdonnell_laser_2004}. This result demonstrates non-classical photon statistics in the collected fluorescence, with the residual $g^{(2)}(0)$ mainly attributed to background counts and finite timing resolution.

\section{Conclusion}\label{sec13}
We demonstrated a long-working distance, high efficiency ion-photon interface with a custom objective composed of commercial off-the-shelf lenses.
The objective provides a working distance of \qty{26.15}{\milli\meter} and does not necessitate complicated chamber geometry such as a recessed viewport or an in-vacuum optical mount.
We achieved high overall photon detection efficiency of \((1.29\pm0.09)\%\) and the system is resilient against system drift over several hours.
The designed ion-photon interface is compatible with a wider range of ion-trap geometries because of relaxed mechanical constraints, making it readily transferable to other trapped-ion quantum network experiments.

\section*{List of abbreviations}

\begin{description}
    \item[NA] Numerical aperture
    \item[WD] Working distance
    \item[SMF] Single-mode fiber
    \item[CF] ConFlat
    \item[MFD] Mode-field diameter
    \item[RMS] Root-mean-square
    \item[CMOS] Complementary metal-oxide-semiconductor
    \item[PSF] Point-spread function
    \item[CAD] Computer-aided design
    \item[HVAC] Heating, ventilation, and air conditioning

\end{description}

\section*{Declarations}

\subsection*{Availability of data and materials}
The datasets generated and analyzed during the current study are available from the corresponding author on reasonable request.

\subsection*{Competing interests}
The authors declare that they have no competing interests.

\subsection*{Funding}
This work was supported by the National Research Foundation of Korea (NRF) grant funded by the Korea government (MSIT) (2022M3H3A1063074, RS-2023-00302576, RS-2024-00466865) and Institute of Information \& Communications Technology Planning \& Evaluation (IITP) grant funded by the Korean government (MSIT) (RS-2022-II221040). The funders had no role in the design of the study, data collection and analysis, decision to publish, or preparation of the manuscript.

\subsection*{Use of large language models}

A large language model (LLM) was used to assist with language editing and improve the clarity of the manuscript. All scientific content and conclusions were reviewed and verified by the authors.

\subsection*{Authors' contributions}
J.C. designed and performed the experiments, analyzed the data, and drafted the manuscript. I.O., C.I., and G.S. contributed to the experimental setup and revising the manuscript. J.K. supervised the project and revised the manuscript. All authors read and approved the final manuscript.

\subsection*{Acknowledgements}
Not applicable.

\begin{appendix}

\section{Optical Bloch Equation Model for Free-Space Detection Efficiency} \label{secA1}

The free-space detection efficiency \(\eta_{\mathrm{fs}}\) is estimated by comparing the measured fluorescence count rate with the steady-state excited-state population predicted by a three-level optical Bloch equation model.
The ion is modeled as a driven \(\Lambda\)-system consisting of \(\lvert S\rangle \equiv \lvert S_{1/2}\rangle\), \(\lvert P\rangle \equiv \lvert P_{1/2}\rangle\), and \(\lvert D\rangle \equiv \lvert D_{3/2}\rangle\).
The \qty{493}{\nano\meter} cooling laser drives the \(\lvert S\rangle \leftrightarrow \lvert P\rangle\) transition, while the \qty{650}{\nano\meter} repump laser drives the \(\lvert D\rangle \leftrightarrow \lvert P\rangle\) transition.

Using the detuning convention \(\delta=\omega_L-\omega_0\) and taking \(\lvert P\rangle\) as the zero of energy, the Hamiltonian in the rotating frame is written in angular-frequency units as
\begin{equation}
\frac{H}{\hbar}
=
\delta_c\lvert S\rangle\langle S\rvert
+
\delta_r\lvert D\rangle\langle D\rvert
+
\frac{\Omega_c}{2}
\bigl(
\lvert S\rangle\langle P\rvert
+
\lvert P\rangle\langle S\rvert
\bigr)
+
\frac{\Omega_r}{2}
\bigl(
\lvert D\rangle\langle P\rvert
+
\lvert P\rangle\langle D\rvert
\bigr),
\end{equation}
where \(\delta_c\) and \(\delta_r\) are the detunings of the cooling and repump lasers, respectively, and \(\Omega_c\) and \(\Omega_r\) are the corresponding Rabi frequencies.

Dissipative dynamics are included through the Lindblad master equation
\begin{equation}
\dot{\rho}
=
-i
\left[
\frac{H}{\hbar},\rho
\right]
+
\sum_{k\in\{c,r\}}
\left(
C_k\rho C_k^\dagger
-
\frac{1}{2}
\left\{
C_k^\dagger C_k,\rho
\right\}
\right),
\end{equation}
with collapse operators \(C_c=\sqrt{\gamma_c}\lvert S\rangle\langle P\rvert\) and \(C_r=\sqrt{\gamma_r}\lvert D\rangle\langle P\rvert\).
The partial decay rates are \(\gamma_c=2\pi\times\qty{15.2}{\mega\hertz}\) and \(\gamma_r=2\pi\times\qty{4.93}{\mega\hertz}\)~\cite{klose_critically_2002}.
The steady-state excited-state population \(\rho_{PP}\) is obtained from \(\dot{\rho}=0\) for each parameter set \(\{\Omega_c,\Omega_r,\delta_c,\delta_r\}\).

During the measurements, the frequency and power of the repump laser are held fixed.
The cooling-laser power \(P_c\) is swept up to \qty{100}{\micro\watt} at four frequency settings separated by \qty{20}{\mega\hertz}.
Both lasers are frequency stabilized using a wavemeter-based feedback loop.
To account for wavemeter calibration uncertainty, the cooling-laser detunings are parameterized as \(\delta_{c,i}=\delta_{c,0}+2\pi\times\{0,-20,-40,-60\}~\mathrm{MHz}\), where the common offset \(\delta_{c,0}\) is determined by the global fit.

The cooling-laser Rabi frequency is modeled as \(\Omega_c(P_c)=\Omega_{c,0}\sqrt{P_c/\qty{1}{\micro\watt}}\), where \(\Omega_{c,0}\) is a fitted scaling parameter.
The four power-scan datasets are fitted simultaneously using
\begin{equation}
R_i(P_c)
=
\eta_{\mathrm{fs}}\gamma_c\,
\rho_{PP}
\bigl(
\Omega_c(P_c),\Omega_r,\delta_{c,i},\delta_r
\bigr)
+
B_{\mathrm{eff}},
\end{equation}
where \(B_{\mathrm{eff}}\) is the effective background count rate, independently measured to be \qty{1.5}{\kilo\hertz} and held fixed in the fit.
The fitted parameters are \(\eta_{\mathrm{fs}}\), \(\delta_{c,0}\), \(\delta_r\), \(\Omega_{c,0}\), and \(\Omega_r\).
Although the repump-laser frequency and power are held fixed, \(\delta_r\) and \(\Omega_r\) are included as fitted parameters because they are not independently calibrated.

\begin{figure}[htbp]
\centering
\includegraphics[width=\linewidth]{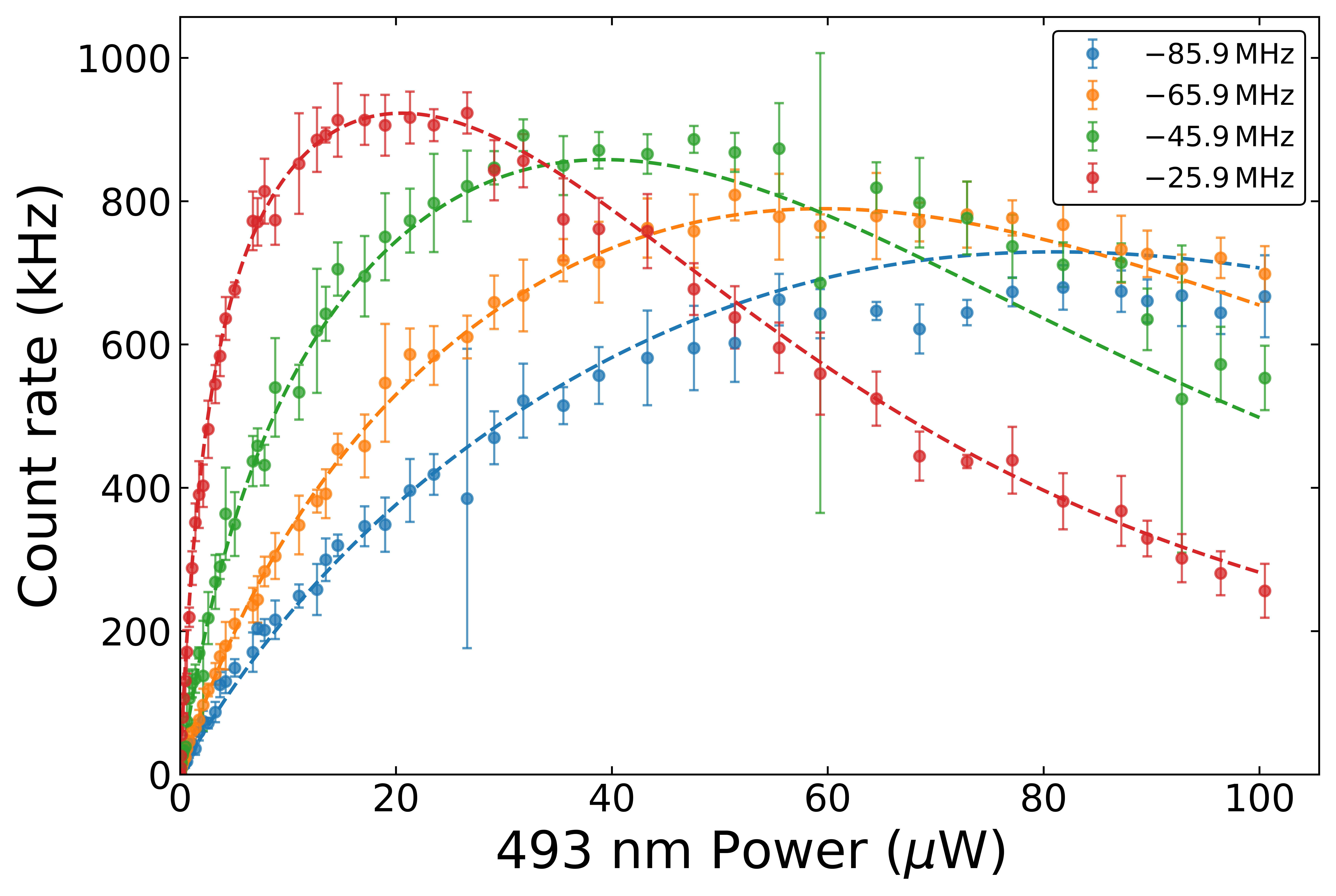}
\caption{
Measured \qty{493}{\nano\meter} fluorescence count rate as a function of the cooling-laser power \(P_c\) for four frequency settings separated by \qty{20}{\mega\hertz}.
Points show the experimental data, and dashed lines show the simultaneous global fit of the steady-state three-level optical Bloch equation model.
The legend indicates the fitted cooling-laser detuning \(\delta_{c,i}/2\pi\) for each dataset.}
\label{figA1}
\end{figure}

The measured fluorescence count rates and the corresponding global fit are shown in Fig.~\ref{figA1}.
The fitted parameters are summarized in Table~\ref{tab:obe_fit}, where the uncertainties correspond to one standard deviation (\(1\sigma\)).
The global fit yields a reduced chi-squared value of \(\chi_\nu^2=1.341\).
The fitted free-space detection efficiency is \(\eta_{\mathrm{fs}}=(3.09\pm0.04)\%\).

\begin{table}[htbp]
\centering
\caption{
Global fit parameters of the steady-state three-level optical Bloch equation model.
The uncertainties represent one-standard-deviation (\(1\sigma\)) fit uncertainties.}
\label{tab:obe_fit}
\begin{tabular}{lll}
\hline
Parameter & Description & Fit result \\
\hline
\(\delta_{c,0}/2\pi\)
& Cooling-laser detuning offset
& \(-25.9 \pm 1.6~\mathrm{MHz}\) \\
\(\delta_r/2\pi\)
& Repump-laser detuning
& \(+23.5 \pm 1.5~\mathrm{MHz}\) \\
\(\Omega_{c,0}/2\pi\)
& Cooling-laser Rabi frequency at \(P_c=\qty{1}{\micro\watt}\)
& \(14.0 \pm 0.3~\mathrm{MHz}\) \\
\(\Omega_r/2\pi\)
& Repump-laser Rabi frequency
& \(31.8 \pm 2.1~\mathrm{MHz}\) \\
\(\eta_{\mathrm{fs}}\)
& Free-space detection efficiency
& \((3.09 \pm 0.04)\%\) \\
\hline
\end{tabular}
\end{table}

\end{appendix}


\begin{thebibliography}{10}
\providecommand{\doi}[1]{\url{https://doi.org/#1}}
\bibcommenthead

\bibitem[\protect\citeauthoryear{Ballance
  et~al.}{2016}]{ballance_high-fidelity_2016}
Ballance CJ, Harty TP, Linke NM, Sepiol MA, Lucas DM.
\newblock High-{Fidelity} {Quantum} {Logic} {Gates} {Using} {Trapped}-{Ion}
  {Hyperfine} {Qubits}.
\newblock Physical Review Letters. 2016 Aug;117(6):060504.
\newblock \doi{10.1103/PhysRevLett.117.060504}.

\bibitem[\protect\citeauthoryear{Gaebler
  et~al.}{2016}]{gaebler_high-fidelity_2016}
Gaebler JP, Tan TR, Lin Y, Wan Y, Bowler R, Keith AC, et~al.
\newblock High-{Fidelity} {Universal} {Gate} {Set} for 9-{Be}+ {Ion} {Qubits}.
\newblock Physical Review Letters. 2016 Aug;117(6):060505.
\newblock \doi{10.1103/PhysRevLett.117.060505}.

\bibitem[\protect\citeauthoryear{Wang et~al.}{2021}]{wang_single_2021}
Wang P, Luan CY, Qiao M, Um M, Zhang J, Wang Y, et~al.
\newblock Single ion qubit with estimated coherence time exceeding one hour.
\newblock Nature Communications. 2021 Jan;12(1):233.
\newblock \doi{10.1038/s41467-020-20330-w}.

\bibitem[\protect\citeauthoryear{Duan and Monroe}{2010}]{duan_colloquium_2010}
Duan LM, Monroe C.
\newblock \textit{{Colloquium}} : {Quantum} networks with trapped ions.
\newblock Reviews of Modern Physics. 2010 Apr;82(2):1209--1224.
\newblock \doi{10.1103/RevModPhys.82.1209}.

\bibitem[\protect\citeauthoryear{Monroe et~al.}{2014}]{monroe_large_2014}
Monroe C, Raussendorf R, Ruthven A, Brown KR, Maunz P, Duan LM, et~al.
\newblock Large {Scale} {Modular} {Quantum} {Computer} {Architecture} with
  {Atomic} {Memory} and {Photonic} {Interconnects}.
\newblock Physical Review A. 2014 Feb;89(2):022317.
\newblock ArXiv:1208.0391 [quant-ph]. \doi{10.1103/PhysRevA.89.022317}.

\bibitem[\protect\citeauthoryear{Hucul et~al.}{2015}]{hucul_modular_2015}
Hucul D, Inlek IV, Vittorini G, Crocker C, Debnath S, Clark SM, et~al.
\newblock Modular entanglement of atomic qubits using photons and phonons.
\newblock Nature Physics. 2015 Jan;11(1):37--42.
\newblock \doi{10.1038/nphys3150}.

\bibitem[\protect\citeauthoryear{Brown et~al.}{2016}]{brown_co-designing_2016}
Brown KR, Kim J, Monroe C.
\newblock Co-designing a scalable quantum computer with trapped atomic ions.
\newblock npj Quantum Information. 2016 Nov;2(1):16034.
\newblock \doi{10.1038/npjqi.2016.34}.

\bibitem[\protect\citeauthoryear{Main et~al.}{2025}]{main_distributed_2025}
Main D, Drmota P, Nadlinger DP, Ainley EM, Agrawal A, Nichol BC, et~al.
\newblock Distributed quantum computing across an optical network link.
\newblock Nature. 2025 Feb;638(8050):383--388.
\newblock \doi{10.1038/s41586-024-08404-x}.

\bibitem[\protect\citeauthoryear{Blinov et~al.}{2004}]{blinov_observation_2004}
Blinov BB, Moehring DL, Duan LM, Monroe C.
\newblock Observation of entanglement between a single trapped atom and a
  single photon.
\newblock Nature. 2004 Mar;428(6979):153--157.
\newblock \doi{10.1038/nature02377}.

\bibitem[\protect\citeauthoryear{Simon and Irvine}{2003}]{simon_robust_2003}
Simon C, Irvine WTM.
\newblock Robust {Long}-{Distance} {Entanglement} and a {Loophole}-{Free}
  {Bell} {Test} with {Ions} and {Photons}.
\newblock Physical Review Letters. 2003 Sep;91(11):110405.
\newblock \doi{10.1103/PhysRevLett.91.110405}.

\bibitem[\protect\citeauthoryear{Moehring
  et~al.}{2007}]{moehring_entanglement_2007}
Moehring DL, Maunz P, Olmschenk S, Younge KC, Matsukevich DN, Duan LM, et~al.
\newblock Entanglement of single-atom quantum bits at a distance.
\newblock Nature. 2007 Sep;449(7158):68--71.
\newblock \doi{10.1038/nature06118}.

\bibitem[\protect\citeauthoryear{Stephenson
  et~al.}{2020}]{stephenson_high-rate_2020}
Stephenson LJ, Nadlinger DP, Nichol BC, An S, Drmota P, Ballance TG, et~al.
\newblock High-{Rate}, {High}-{Fidelity} {Entanglement} of {Qubits} {Across} an
  {Elementary} {Quantum} {Network}.
\newblock Physical Review Letters. 2020 Mar;124(11):110501.
\newblock \doi{10.1103/PhysRevLett.124.110501}.

\bibitem[\protect\citeauthoryear{Olmschenk
  et~al.}{2009}]{olmschenk_quantum_2009}
Olmschenk S, Matsukevich DN, Maunz P, Hayes D, Duan LM, Monroe C.
\newblock Quantum {Teleportation} {Between} {Distant} {Matter} {Qubits}.
\newblock Science. 2009 Jan;323(5913):486--489.
\newblock \doi{10.1126/science.1167209}.

\bibitem[\protect\citeauthoryear{O’Reilly et~al.}{2024}]{oreilly_fast_2024}
O’Reilly J, Toh G, Goetting I, Saha S, Shalaev M, Carter AL, et~al.
\newblock Fast {Photon}-{Mediated} {Entanglement} of {Continuously} {Cooled}
  {Trapped} {Ions} for {Quantum} {Networking}.
\newblock Physical Review Letters. 2024 Aug;133(9):090802.
\newblock \doi{10.1103/PhysRevLett.133.090802}.

\bibitem[\protect\citeauthoryear{Kim et~al.}{2011}]{kim_efficient_2011}
Kim T, Maunz P, Kim J.
\newblock Efficient collection of single photons emitted from a trapped ion
  into a single-mode fiber for scalable quantum-information processing.
\newblock Physical Review A. 2011 Dec;84(6):063423.
\newblock \doi{10.1103/PhysRevA.84.063423}.

\bibitem[\protect\citeauthoryear{Myerson
  et~al.}{2008}]{myerson_high-fidelity_2008}
Myerson AH, Szwer DJ, Webster SC, Allcock DTC, Curtis MJ, Imreh G, et~al.
\newblock High-{Fidelity} {Readout} of {Trapped}-{Ion} {Qubits}.
\newblock Physical Review Letters. 2008 May;100(20):200502.
\newblock \doi{10.1103/PhysRevLett.100.200502}.

\bibitem[\protect\citeauthoryear{Noek et~al.}{2013}]{noek_high_2013}
Noek R, Vrijsen G, Gaultney D, Mount E, Kim T, Maunz P, et~al.
\newblock High speed, high fidelity detection of an atomic hyperfine qubit.
\newblock Optics Letters, Vol 38, Issue 22, pp 4735-4738. 2013
  Nov;\doi{10.1364/OL.38.004735}.

\bibitem[\protect\citeauthoryear{Carter et~al.}{2024}]{carter_ion_2024}
Carter AL, O’Reilly J, Toh G, Saha S, Shalaev M, Goetting I, et~al.
\newblock Ion trap with in-vacuum high numerical aperture imaging for a
  dual-species modular quantum computer.
\newblock Review of Scientific Instruments. 2024 Mar;95(3):033201.
\newblock \doi{10.1063/5.0180732}.

\bibitem[\protect\citeauthoryear{Saha et~al.}{2025}]{saha_high-fidelity_2025}
Saha S, Shalaev M, O’Reilly J, Goetting I, Toh G, Kalakuntla A, et~al.
\newblock High-fidelity remote entanglement of trapped atoms mediated by
  time-bin photons.
\newblock Nature Communications. 2025 Mar;16(1):2533.
\newblock \doi{10.1038/s41467-025-57557-4}.

\bibitem[\protect\citeauthoryear{Maunz et~al.}{2007}]{maunz_quantum_2007}
Maunz P, Moehring DL, Olmschenk S, Younge KC, Matsukevich DN, Monroe C.
\newblock Quantum interference of photon pairs from two remote trapped atomic
  ions.
\newblock Nature Physics. 2007 Aug;3(8):538--541.
\newblock \doi{10.1038/nphys644}.

\bibitem[\protect\citeauthoryear{Crain et~al.}{2014}]{crain_individual_2014}
Crain S, Mount E, Baek S, Kim J.
\newblock Individual addressing of trapped {171Yb}+ ion qubits using a
  microelectromechanical systems-based beam steering system.
\newblock Applied Physics Letters. 2014 Jan;105(18):181115.
\newblock \doi{10.1063/1.4900754}.

\bibitem[\protect\citeauthoryear{Pritchard et~al.}{2016}]{pritchard_long_2016}
Pritchard JD, Isaacs JA, Saffman M.
\newblock Long working distance objective lenses for single atom trapping and
  imaging.
\newblock Review of Scientific Instruments. 2016 Jul;87(7):073107.
\newblock \doi{10.1063/1.4959775}.

\bibitem[\protect\citeauthoryear{Li
  et~al.}{2020}]{li_high-numerical-aperture_2020}
Li S, Li G, Wu W, Fan Q, Tian Y, Yang P, et~al.
\newblock High-numerical-aperture and long-working-distance objective for
  single-atom experiments.
\newblock Review of Scientific Instruments. 2020 Apr;91(4):043104.
\newblock \doi{10.1063/5.0001637}.

\bibitem[\protect\citeauthoryear{Schupp et~al.}{2021}]{schupp_interface_2021}
Schupp J, Krcmarsky V, Krutyanskiy V, Meraner M, Northup TE, Lanyon BP.
\newblock Interface between {Trapped}-{Ion} {Qubits} and {Traveling} {Photons}
  with {Close}-to-{Optimal} {Efficiency}.
\newblock PRX Quantum. 2021 Jun;2(2):020331.
\newblock \doi{10.1103/PRXQuantum.2.020331}.

\bibitem[\protect\citeauthoryear{Hanbury~Brown and
  Twiss}{1956}]{hanbury_brown_test_1956}
Hanbury~Brown R, Twiss RQ.
\newblock A {Test} of a {New} {Type} of {Stellar} {Interferometer} on {Sirius}.
\newblock Nature. 1956 Nov;178(4541):1046--1048.
\newblock \doi{10.1038/1781046a0}.

\bibitem[\protect\citeauthoryear{Diedrich and
  Walther}{1987}]{diedrich_nonclassical_1987}
Diedrich F, Walther H.
\newblock Nonclassical radiation of a single stored ion.
\newblock Physical Review Letters. 1987 Jan;58(3):203--206.
\newblock \doi{10.1103/PhysRevLett.58.203}.

\bibitem[\protect\citeauthoryear{McDonnell et~al.}{2004}]{mcdonnell_laser_2004}
McDonnell MJ, Stacey DN, Steane AM.
\newblock Laser linewidth effects in quantum state discrimination by
  electromagnetically induced transparency.
\newblock Physical Review A. 2004 Nov;70(5):053802.
\newblock \doi{10.1103/PhysRevA.70.053802}.

\bibitem[\protect\citeauthoryear{Klose et~al.}{2002}]{klose_critically_2002}
Klose JZ, Fuhr JR, Wiese WL.
\newblock Critically {Evaluated} {Atomic} {Transition} {Probabilities} for {Ba}
  {I} and {Ba} {II}.
\newblock Journal of Physical and Chemical Reference Data. 2002
  Mar;31(1):217--230.
\newblock \doi{10.1063/1.1448482}.

\end{thebibliography}
\end{document}